\documentclass[nofootinbib]{revtex4}

\usepackage{graphicx}
\usepackage{dcolumn}
\usepackage{bm}
\usepackage{nameref}
\usepackage{hyperref}
\usepackage[utf8]{inputenc}
\usepackage{mathtools}
\usepackage{siunitx}
\usepackage{amssymb}
\usepackage[T1]{fontenc}
\usepackage{mathptmx}
\usepackage{upgreek}
\usepackage{amsmath}
\usepackage[version=3]{mhchem}
\usepackage[thinc]{esdiff}
\usepackage{derivative}
\usepackage{etoolbox}
\usepackage{xcolor}
\usepackage{mathrsfs}
\usepackage{lipsum}
\usepackage{rsfso}
\usepackage{orcidlink}
\makeatletter

\def\@email#1#2{%
 \endgroup
 \patchcmd{\titleblock@produce}
  {\frontmatter@RRAPformat}
  {\frontmatter@RRAPformat{\produce@RRAP{*#1\href{mailto:#2}{#2}}}\frontmatter@RRAPformat}
  {}{}
}%
\makeatother
\begin{document}

\title{Quantum phase transitions of Dirac particles affected from magnetized 2+1 curved background}

\author{N. Sahan}
\orcidlink{{0000-0002-8321-0753}}

\affiliation{Department of Physics, Faculty of Science, Akdeniz University, 07058 Antalya, Turkey}
\author{E. Sucu}
\orcidlink{0009-0000-3619-1492}

\affiliation{Department of Physics, Eastern Mediterranean
University, Famagusta, 99628 North Cyprus via Mersin 10, Turkey}

\date{\today}

\date{\today}

\begin{abstract}

In this research,  we investigate the quantum and classical phase transitions of the Dirac particles in a homogeneously magnetized curved rotating 2+1 dimensional spacetime. We consider the intricate relationship between geometry and quantum phase events through the study of quantum electrodynamics in the rotating curved spacetime. Using methods from quantum electrodynamics and statistical mechanics, the study examines the effects of an external magnetic field, the background rotation parameter, and curvature on the characteristics of quantum and classical phase transitions, focusing on critical points and scaling behavior, and we see that as thermal fluctuations get closer to zero, quantum fluctuations begin to dominate at this system.

\bigskip

\end{abstract}
\maketitle

\section{\label{sec:level1}Introduction}

Theoretical physicists have long been interested in investigating the interaction between unusual spacetime dimensions and quantum phenomena to understand the principles of the quantum realm and the structure of spacetime. Especially, 2+1 dimensional spacetime satisfies many simplifications in general relativity \cite{carlip2003quantum,ba1992nados,andrade20052+}, quantum electrodynamics \cite{mross2016explicit,sucu2007exact}, and a lot of applications based on their merging \cite{gecim2017dirac,gecim2019quantum,he20172+,gecim2017gup,babar2020effect,gecim2018quantumm,mazharimousavi2014new,gecim2018quantummm,tekincay2021exotic,dernek2018relativistic,gecim2017diraccc,gecim2013tunnelling,maluf2022new,gecim2015dirac}. Additionally, in materials science, interest in spatially 2-dimensional (monolayer) materials is increasing day by day and their physical properties are better understood.\cite{gibertini2019magnetic,qiu2021visualizing,skinner2013effect}. The idea of a 2+1 rotating curved background is an exciting line of inquiry that provides an interesting divergence from standard models of spacetime curvature. In this connection, the 2+1 dimensional rotating curved backgrounds are considered in different ways, with quantum electrodynamics, general relativity, and classical mechanics\cite{adamo2022classical,giataganas2021entropy}. Essentially, this paradigm provides a distinct viewpoint for examining how physical systems or elementary particles interact with such a spacetime background. Also, Investigating quantum phenomena within the framework of rotating curved backgrounds offers a rich environment for studying the interaction between quantum and classical physics. The interface between quantum phase transitions and 2+1 rotating curved backgrounds, where emergent phenomena of great importance arise from the dynamics of spacetime curvature and quantum fluctuations, is one especially interesting field of study. Therefore, it is very important to understand quantum phase transitions by examining the energy eigenvalues of particles in such a system and writing the partition function of this system. In this connection, to investigate the physical properties of a curved background, it has generally been studied classical phase transitions stemming from thermal fluctuations. Therefore, in the study, we consider the quantum phase transitions of the magnetized, rotating 2+1 dimensional curved background in the limit where thermal fluctuations go to absolute temperature zero and the classical phase transitions based on thermal fluctuations.  

Our understanding of the universe is based on quantum statistics, general relativity, and quantum electrodynamics (QED) \cite{berestetskii1982quantum,gell1954quantum}, which provide insights into the behavior of matter and energy at various scales. In this wide field of science, research on quantum phase transitions (QPTs)\cite{zurek2005dynamics,greentree2006quantum,hartnoll2007theory,dvali2014black,sachdev1999quantum,vojta2003quantum,carollo2020geometry,kol2006phase}, stands out as a very fascinating and important field. absolute temperature zero, quantum phase transitions take place, propelled by quantum fluctuations instead of thermal energy. These fluctuations indicate a fundamental shift in a system's ground state brought about by changes in characteristics like particle density, pressure, or magnetic field. Quantum phase transitions denote sudden variations in a quantum system's ground state in response to outside factors like magnetic field intensity or pressure. These transitions take place at absolute temperature zero and are typified by the appearance of new quantum states and collective behaviors that frequently go against conventional wisdom.

The remainder of the paper is organized as follows. In the current section 2, we review some basic information about the 2+1 rotating curved background and its fundamental properties. In section 3 we find energy eigenvalues that Dirac particles in 2+1 dimensional rotated curved background with constant magnetic field. Section 4, the calculation of the partition function of this system. Furthermore, we derive thermodynamic properties and magnetization.

\section{Physical properties of rotating curved background}

This section provides a concise overview of the spacetime of the 2+1 rotating curved background, which was first described by Y. Nutku \cite{nutku1993exact}. Topologically massive gravity equations with a cosmological constant are given as follows

\begin{equation}
    G^{\mu}_{\nu}+ \frac{\eta^{\mu \beta \alpha}}{\mu}\nabla_{\alpha}(R_{\nu \beta}-\frac{1}{4}R g_{\nu \beta})= \lambda \delta_{\nu}^{\mu}
\end{equation}

in which the Einstein and Ricci tensors are represented by $G_{\mu \nu}$ and $R_{\mu \nu}$, respectively. In three dimensions, $\eta^{\mu \beta \alpha}$ represents the Levi-Civita tensor, while $\mu$ and $\lambda$ stand for the Deser-Jackiw-Templeton (DJT) parameter and the cosmological constant, respectively. The following are the equations of Einstein for a perfect fluid source are given

\begin{equation}
    G_{\mu \nu}=T_{\mu \nu}
\end{equation}
in which 

\begin{equation}
    T_{\mu \nu}=p g_{\mu \nu}+ (p+\rho)u_{\mu} u_{\nu}
\end{equation}

 $u_{\mu}$ is the fluid's timelike unit 4-velocity, while p and $\rho$ stand for the fluid's pressure and mass density, respectively. The generic metric expression of the Perfect fluid sources in 2+1 dimensions spacetime can be defined as follows \cite{nutku1993exact}

\begin{equation}
ds^2= \left(dt - \Omega(r) d\phi\right)^2-dr^2 - r^2d\phi^2 .
\label{0}
\end{equation}
in which
\begin{equation}
    \Omega(r)=-\frac{kr^2}{2}
\end{equation}
where k is the vorticity parameter characterizing the spacetime and keep in mind that the Gurses metrics described in \cite{gurses1994perfect} are a special instance of the research metric \eqref{0}. The characteristics for three-dimensional metrics are also examined in more detail \cite{gleiser2006closed,gurses2010godel}. The metric itself has signatures of (+, -, -).

\section{Dirac particles in 2+1 dimensional rotated curved background with
homogenous magnetic field}

To investigate the physical behavior of Dirac particles in a homogeneously magnetized 2+1 dimensional rotated curved space-time, we consider the metric as emphasized in \cite{nutku1993exact,gurses1994perfect, torovs2022generation}: 
  
\begin{equation}
ds^2= \left(dt + \Omega(r)d\phi\right)^2-dr^2 - r^2d\phi^2 .
\label{1}
\end{equation}

The contravariant metric tensor $g^{\mu\nu}$ ,associated with metric, is in the following manner:

\begin{equation}
g^{\mu\nu} = \begin{pmatrix}
1- \frac{\Omega^2(r)}{r^2} & 0 & \frac{\Omega(r)}{r^2} \\ 
0 & -1 & 0 \\ 
\frac{\Omega (r)}{r^2} & 0 & -\frac{1}{r^2}
\end{pmatrix}.
\label{2}
\end{equation}

Additionally, the triads, denoted as  $e^{a\nu}$ and related by $e^{a\mu}e^{\phantom{a}\nu}_a=g^{\mu\nu}$, are expressed by

\begin{equation}
    e^{a\nu} =  \begin{pmatrix}
1 & 0 & 0 \\ 
0 & -1 & 0 \\ 
 \frac{\Omega (r)}{r} & 0 & -\frac{1}{r}
\end{pmatrix}.
\label{3}
\end{equation}

In the context of 2+1 gravity, the Dirac equation is extensively studied \cite{sucu2004dirac,sakalli2004solution,dogan2019quasinormal} and it is expressed as follows \cite{dirac1928quantum,thaller2013dirac}

\begin{equation}
\{i\hbar\bar{\sigma}^\mu(x)[\partial_\mu - \Gamma_\mu(x) + ieA_\mu]\}\Psi(x) = m_ec^2\Psi(x),
\label{4}
\end{equation}

where e and $m_e$ respectively signify the charge and mass of the electron. $A_\mu$ denotes the components of the electromagnetic potential, expressed as $A_\mu = \left(0, 0, A_2\right)$ in which $A_2 = -Br^2/2$ with B representing the magnetic field. $\tilde{\sigma}^a = (\sigma^3,i\sigma^1,i\sigma^2)$ with $\sigma^1, \sigma^2, \sigma^3$ being Pauli matrices. The spin connection components are symbolized by $\Gamma_\mu$ and, their computations for this metric are carried out as illustrated

\begin{eqnarray}
\Gamma_0 &=& -\frac{\Omega^\prime(r)}{8r}[\sigma^2,\sigma^1],
 \nonumber\\
\Gamma_1 &=& \frac{i\Omega^\prime(r)}{8r}[\sigma^2,\sigma^3], \nonumber \\
\Gamma_2 &=& \frac{1}{4}\Biggl\{\left(1 - \frac{\Omega(r)\Omega^\prime(r)}{2r}\right)[\sigma^1,\sigma^2] + \frac{i\Omega^\prime(r)}{2}[\sigma^3,\sigma^1]\Biggr\}.
\label{5}
\end{eqnarray}

In this case, the Dirac equation becomes

\begin{equation}
\left[-i\sigma^3\partial_t + \sigma^1\left(\partial_r + \frac{1}{2r}\right) + \sigma^2\left(\frac{\partial_\phi}{r} - \frac{\Omega(r)}{r}\partial_t - \frac{ieBr}{2}\right) - \frac{\Omega^\prime(r)}{4r} + m_e\right]\Psi = 0. 
\label{6}
\end{equation}

We can rewrite $\sigma ^{1}$ and $\sigma ^{2}$ matrices in terms of right and left circular spin matrices $\left(\sigma^+ \ \text{and} \ \sigma^-\right)$ in the way indicated

\begin{equation}
     \sigma^\pm = \frac{\sigma^1 \pm i\sigma^2}{2}, 
     \label{7}
\end{equation}
   
Then, we can regroup Eq. (\ref{6}) in terms of $\sigma^\pm$ as

\begin{equation}
   \left( -i\sigma ^3E + \sigma^+ \partial_- + \sigma^- \partial_+ - \frac{\Omega^\prime(r)}{4r} + m_e\right)\Psi = 0,
   \label{8}
\end{equation}

where $\Psi$ is a two-component spinor representing pasitive and negative energy eigen functions, and as the $\phi$ and $t$ are cyclic coordinates, thanks to the separation of variables method we ansatz the $\Psi$ as follows 

\begin{equation}
    \Psi = \frac{e^{i\left(m\phi - Et\right)}}{\sqrt{r}}\chi(r)
    \label{9}
\end{equation}

with $\chi(r)=[\chi_+(r) \quad \chi_-(r)]^\text{T}$ and, the raising and lowering operators for the angular variables $\partial_\pm$\ are

\begin{equation}
   \partial_\pm = \partial_r \pm i\frac{\partial_\phi}{r}\mp\frac{\Omega(r)}{r}E\pm m_e\omega_c r, 
   \label{10}
\end{equation}

where $\omega_c$ represents cyclotron frequency. In this case, Eq. (\ref{6}) becomes as follows

\begin{equation}
    \begin{pmatrix}
-E - \frac{\Omega^{\prime}(r)}{4r} + m_e & \partial_-\\
\partial_+ & E - \frac{\Omega^\prime(r)}{4r} + m_e
\end{pmatrix}\begin{pmatrix}
\chi_+(r)\\
\chi_-(r)
\end{pmatrix} = 0 
\label{11}
\end{equation}

where we choose $\Omega(r) = - kr^2/2$ with k representing the rotating velocity parameter of the background.

\begin{equation}
   \left[\frac{d^2}{dr^2} - \left(\frac{Ek}{2} + m_e\omega_c\right)^2r^2 + \left(Ek + 2m_e \omega_c\right)\left(m - \frac{1}{2}\right) -  \lambda - \frac{m^2}{r^2} -\ \frac{m}{r^2}\right]\chi_\pm(r) = 0,
   \label{12}
\end{equation}

 where $\lambda$ is an eigenvalue which satisfies $\partial_\mp\partial_\pm \chi_\pm = \lambda \chi_\pm$. Therefore,  square-integrable radial spinors are 

\begin{equation}
\chi_\pm(r) = \frac{C_\pm e^{-\frac{\left(Ek +2m_e\omega_c\right)r^2}{4}}}{\sqrt{r}}\left[\left( \frac{Ek + 2m_e\omega_c}{2}\right)r^2\right]^{\frac{m}{2} + \frac{1}{4}}
    {}_1F_1\left(\frac{\lambda}{2Ek + 4m_e\omega_c}, m + \frac{1}{2},  \frac{\left(Ek + 2m_e\omega_c\right)r^2}{2}\right),
    \label{13}
\end{equation}

where $C_\pm$ stand for arbitrary constants and, ${}_1F_1$ is the hypergeometric function. From the polynomial condition of ${}_1F_1$, $\lambda/(2Ek + 4m_e\omega) = -n$ with n being a positive integer, and hence we can determine the energy spectra as

\begin{equation}
   E^\pm = kn \pm \sqrt{k^2n^2 + 4m_e\omega_c n + m_e^2 + \frac{m_ek}{2} + \frac{k^2}{16}}. 
   \label{14}
\end{equation}

Then, the dimensionless energy spectra for the system are given by

\begin{equation}
   \epsilon^\pm = \frac{E^\pm}{m_ec^2} = 2Kn \pm \sqrt{\left(2Kn\right)^2+4\gamma n+\left(1+\frac{K}{2}\right)^2},
   \label{15}
\end{equation}

where $K = \hbar k/2m_ec $ and $\gamma = \hbar \omega_c/ m_e c^2$.

\section{Statistical Mechanics and Thermodynamics Properties}

\subsection{Partition Function}

The partition function is an important tool for transition from the microscopic to the macroscopic system. In this section,  we construct the partition function from the energy eigenvalues in Eq. (\ref{15}) to analyze the quantum and classical phase transitions of the system. However, as it is difficult to make direct calculations with these energy eigenvalues, we simplify the energy eigenvalues under $(1+K/2)^2= \gamma^2/K^2$ which may be written in the following
fashion

\begin{eqnarray}
\epsilon^+ &=& 4K_\gamma n+\frac{\gamma}{K_\gamma}, \label{16}\\
\epsilon^- &=&-\frac{\gamma}{K_\gamma} \label{17},
\end{eqnarray}

where $K_\gamma$ expresses all the roots of K that satisfy the above special condition. In this case, at least,  we can easily determine quantum phase transitions for some values of the k rotating parameter of the curved background. The positive and negative energy solutions determine the partition function for the left and right phases (LP and RP). Since we focus on particle solutions, not antiparticles, in our study we find out a partition function for the right and left phase by changing the direction of the magnetic field in the positive energy eigenvalue \cite{mandal2012noncommutative, frassino2020thermodynamics}. Thus, the partition function can be given by 

\begin{equation}
    Z_{L,R}=\sum_{n_r, n_l}ge^{-\tilde{\beta} \epsilon^+_n}, 
    \label{18}
\end{equation}
where $ n = n_r + \left[\left| n_r - n_l \right| -\left(n_r - n_l\right)\right]/2$, with $n_r$ and $n_l$ representing quantum numbers, g is the density of states and $\tilde{\beta} = m_e c^2 / k_B T,$ with $k_B$ being Boltzmann constant and $T$ signifying absolute temperature. It is possible to establish the degree of degeneracy of each energy level by g. To evaluate g, we divide each chiral phase of the system into two distinct sub-systems based on $n_r$ and $n_l$. If we consider a free electron in a standard commutative space-time and further restrict it to a finite area $L^2$, with L representing the rest-frame length, we can distinguish two cases: one where $n_{r} < n_{l}$, and the other where $n_{r} \ge n_{l}$. The number of energy levels in the region surrounding p$_{x}$ and p$_{y}$, denoted as dp$_{x} $dp$_{y}$ can be expressed as

\begin{eqnarray}
      dg &=& \frac{L^{2}}{2\hbar^2}dp_{x}dp_{y}  = \frac{L^{2}}{2\hbar^2c^2}  EdE d\phi,\nonumber \\
         &=&  \frac{1}{2}dP_xdP_y = \frac{1}{2}\epsilon d\epsilon d\phi, \label{19}
\end{eqnarray}

where $P_x$ and $P_y$ are dimensionless momentum components $\left(P_x = \frac{Lp_x}{\hbar}, P_y = \frac{L p_y}{\hbar}\right)$. Thus, g for $n_r \ge n_l$ under $E_n < E_{f} < E_{n+1}$, where $E_f$ is the free energy of the electron, can be calculated as

\begin{eqnarray}
    g_{n_r \ge n_l} &=&  \frac{1}{2}\int_0^{2\pi}d\phi\int_{\epsilon_n}^{\epsilon_{n+1}}\epsilon d\epsilon \nonumber \\
&=& 4K_\gamma\pi\left[\left(1+2n_r\right)2K_\gamma +\frac{\gamma}{K_\gamma}\right]. \label{20}
\end{eqnarray}

 The relation between $g_{n_r \ge n_l}$ and $g_{n_r < n_l}$ is given by $ g_{n_r < n_l} = g_{n_r \ge n_l}/D$, where D represents degeneracy number and equals to n + 1. Therefore,

\begin{equation}
    g_{n_r < n_l} = \frac{4K_\gamma\pi}{n_l + 1}\left[\left(1+2n_l\right)2K_\gamma +\frac{\gamma}{K_\gamma}\right]. 
    \label{21}
\end{equation}

When we include the zero mode term, the one-particle partition function for the LP yields

\begin{equation}
    Z_L = 4K_\gamma\pi \Biggl\{\left(2K_\gamma+\frac{\gamma}{K_\gamma}\right)e^{-\frac{\tilde{\beta}\gamma}{K_\gamma}} + \sum_{n_r = 1}^\infty\left[\left(1+2n_r\right)2K_\gamma +\frac{\gamma}{K_\gamma}\right]e^{-\tilde{\beta}\left(4K_\gamma n_r+\frac{\gamma}{K_\gamma}\right)} + \sum_{n_l = 1}^\infty \frac{\left[\left(1+2n_l\right)2K_\gamma +\frac{\gamma}{K_\gamma}\right]}{n_l + 1}e^{-\tilde{\beta}\left(4K_\gamma n_l+\frac{\gamma}{K_\gamma}\right)}\Biggr\}.
    \label{22}
\end{equation}

Under the condition $n_l - n_r = n_r^\prime - n_l^\prime$ and $n_r + n_r^\prime - n_l^\prime = n^\prime$, the second and third sum terms will give the equivalent results. Thus, Eq. (\ref{22}) becomes

\begin{equation}
    Z_L = 4K_\gamma\pi  \Biggl\{\left(2K_\gamma+\frac{\gamma}{K_\gamma}\right)e^{-\frac{\tilde{\beta}\gamma}{K_\gamma}} + 2\sum_{n_r = 1}^\infty\left[\left(1+2n_r\right)2K_\gamma +\frac{\gamma}{K_\gamma}\right]e^{-\tilde{\beta}\left(4K_\gamma n_r+\frac{\gamma}{K_\gamma}\right)}\Biggr\}.
    \label{23}
\end{equation}
  
We refer to the sum term as $s_{n_r}$ and it can be clearly seen that it is a geometric series
\begin{eqnarray}
    s_{n_r} &=& 2X^{\frac{\gamma}{4K_\gamma^2}}\sum_{n_r = 1}^\infty\left[\left(1+ 2n_r\right)2K_\gamma + \frac{\gamma}{K_\gamma}\right]X^{n_r}, \nonumber\\
     &=&  \frac{2X^{\frac{\gamma}{4K_\gamma^2} + 1}}{1 - X}\left(\frac{4K_\gamma}{1 - X} + 2K_\gamma+\frac{\gamma}{K_\gamma}\right). 
     \label{24}
\end{eqnarray}

where $X = e^{-4\tilde{\beta} K_\gamma}$. Once substituting Eq. (\ref{24}) into Eq. (\ref{23}), the one-particle partition function for the LP can be determined as
\begin{eqnarray}
    Z_L &=& 4K_\gamma\pi X^{\frac{\gamma}{4K_\gamma^2}}\left[\left( 2K_\gamma + \frac{\gamma}{K_\gamma}\right)\left(\frac{1 + X}{1 - X}\right) + \frac{8K_\gamma X}{\left(1 - X\right)^2}\right], \nonumber \\
    &=& 4K_\gamma\pi e^{-\frac{\tilde{\beta}\gamma}{K_\gamma}}\left[\left( 2K_\gamma + \frac{\gamma}{K_\gamma}\right)\coth\left(2\tilde{\beta}K_\gamma\right) + \frac{2K_\gamma}{\sinh^2\left(2\tilde{\beta}K_\gamma\right)}\right].
    \label{25}
\end{eqnarray}

By performing analogous procedures changing the direction of the magnetic field $\left(\gamma \rightarrow \gamma_{\_} = -\gamma\right)$ without the zero mode term, it can be readily calculated that the one-particle partition function for the RP is as stated below

\begin{equation}
    Z_R = 4K_{\gamma_{\_}}\pi e^{-\frac{\tilde{\beta}\gamma{\_}}{K_{\gamma_{\_}}}}\left[\left( 2K_{\gamma_{\_}} + \frac{\gamma_{\_}}{K_{\gamma_{\_}}}\right)\text{csch}\left(2\tilde{\beta}K_{\gamma_{\_}}\right) + \frac{2K_{\gamma_{\_}}}{\sinh^2\left(2\tilde{\beta}K_{\gamma_{\_}}\right)}\right].
    \label{26}
\end{equation}

\subsection{ Magnetization}
The magnetization $\left(M\right)$ is a useful thermodynamic variable to examine quantum phase transitions. To investigate the magnetic behaviour of homogeneously magnetized Dirac particles, the dimensionless magnetization $\left(\tilde{M}\right)$ at finite temperature, which can be defined as the ratio of magnetization to Bohr magneton $\left(M/\mu_B\right)$, can be calculated by

\begin{equation}
    \tilde{M} =  \frac{1}{\tilde{\beta}}\frac{\partial \ln Z}{\partial \gamma}.
    \label{27}
\end{equation}

\begin{figure}[hbt!]
\centering
\begin{minipage}{.2\linewidth}
  \centering
  \includegraphics[width=\linewidth]{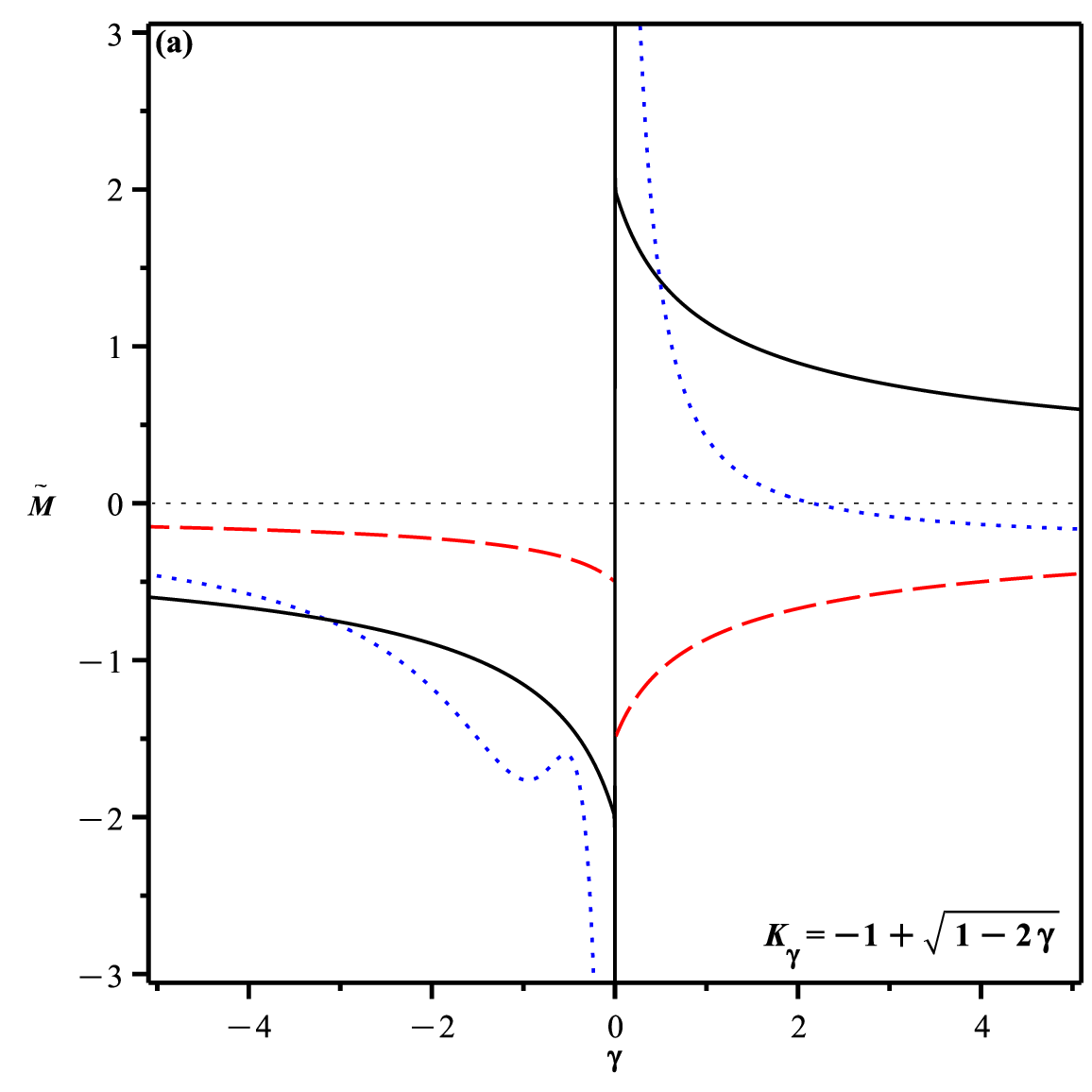}
  \label{Mag1}
\end{minipage}
\quad
\begin{minipage}{.2\linewidth}
  \centering
  \includegraphics[width=\linewidth]{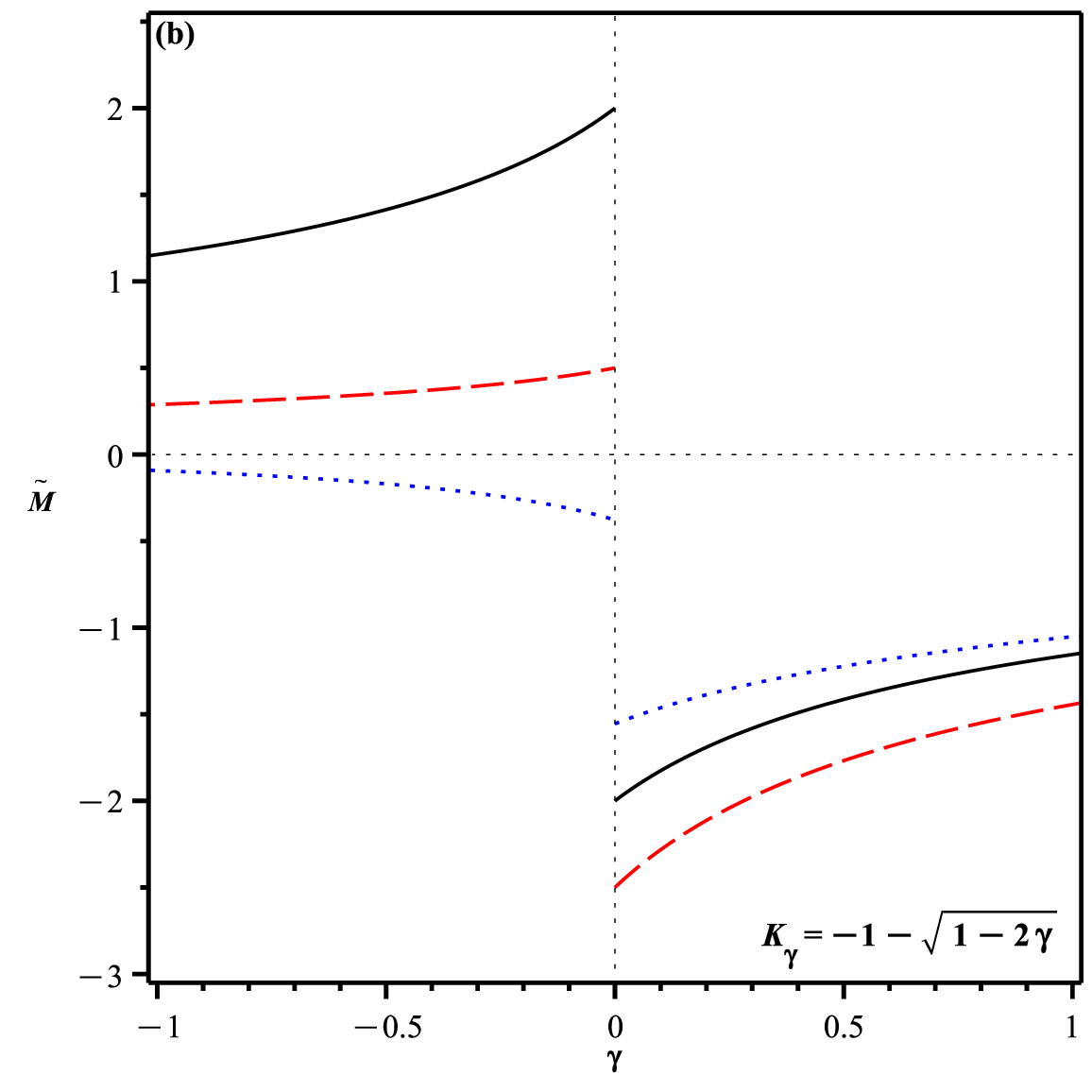}
  \label{Mag2}
\end{minipage}
\begin{minipage}{.2\linewidth}
  \centering
  \includegraphics[width=\linewidth]{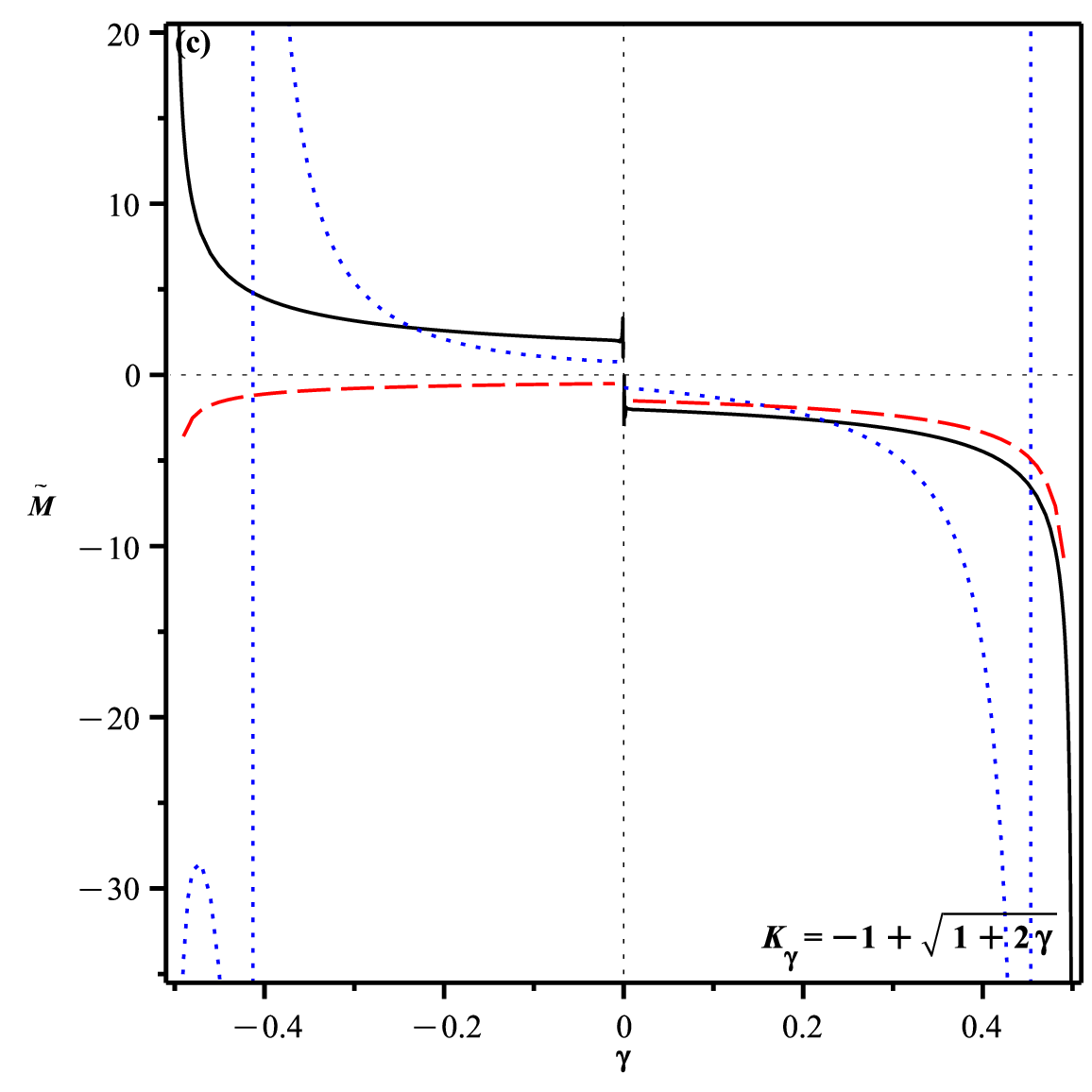}
  \label{Mag3}
\end{minipage}
\quad
\begin{minipage}{.2\linewidth}
  \centering
  \includegraphics[width=\linewidth]{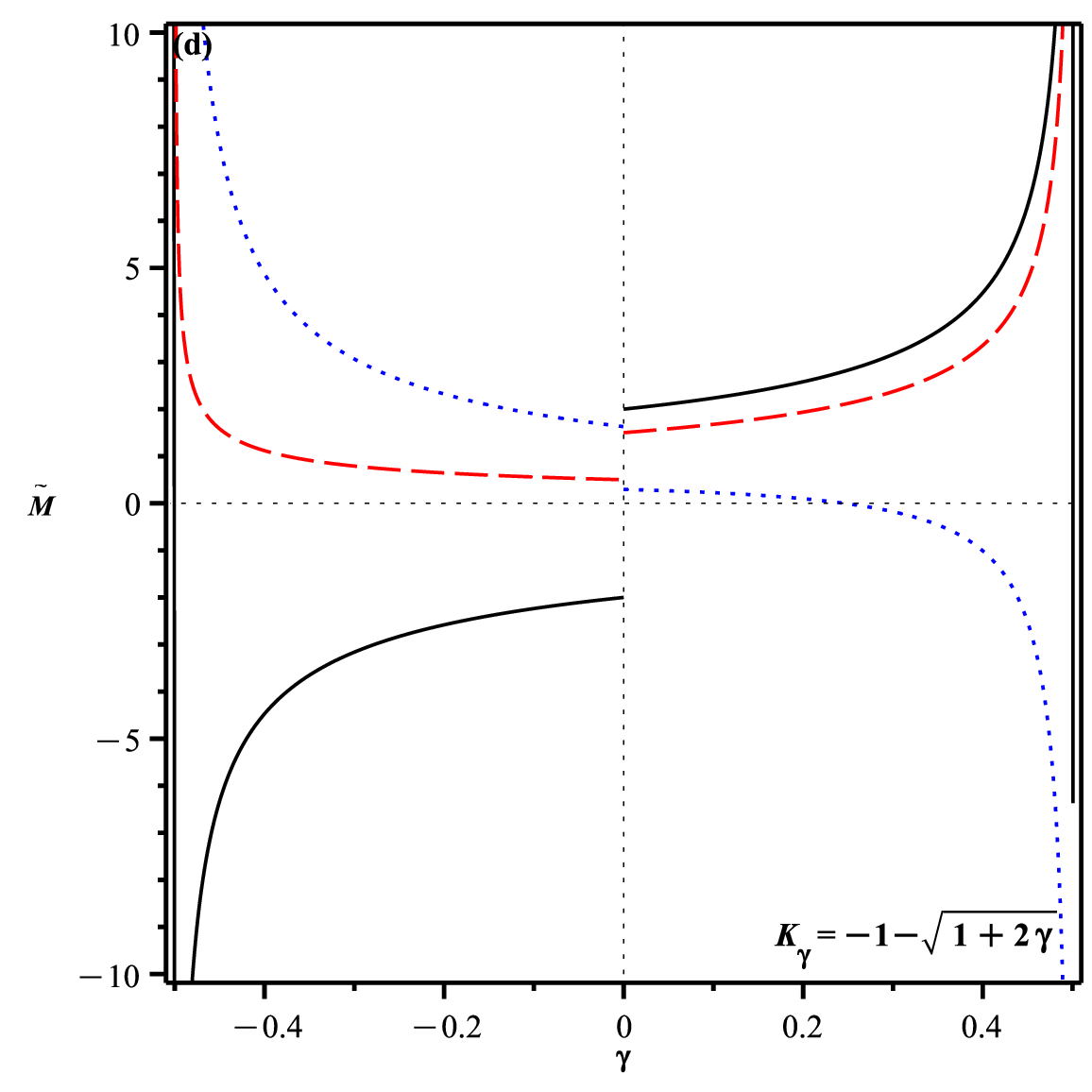}
  \label{Mag4}
\end{minipage}
\caption{(color online). Plots of scaled magnetization versus $\gamma$ for all roots of K in different values of the inverse temperature, $\tilde{\beta} = 10^{-10}$ (solid), $1$ (dotted), $10^{10}$ (dashed), respectively.} 
\label{Fig1}
\end{figure}

One numerically evaluates and is illustrated in Fig. 1 for all K roots.  M manifests a discontinuity at the finite temperature at the $\gamma_c = 0$ where $\gamma_c$ is the critical magnetic field.

$\Tilde{M}$ has $\gamma$ dependence shown in Fig. \ref{Fig1} for various values of $K_\gamma$ corresponding to case (a), (b), (c) and (d). Plots with Right phase (RP) curves on the right side (\(\gamma > 0\)) and Left phase (LP) curves on the left are depicted, respectively. Generally speaking, \(\tilde{M}\) behaves differently for positive and negative \(\gamma\). In (a), for example, \(\tilde{M}\) fluctuates quickly for negative \(\gamma\) and stabilises for positive \(\gamma\). This is because \(K_{\gamma} = -1 + \sqrt{1 - 2\gamma}\)). A smoother transition can be seen in (b) (\(K_{\gamma} = -1 - \sqrt{1 - 2\gamma}\)), \(\tilde{M}\). For negative \(\gamma\), which are not physical for \(|\gamma| > 0.5\), (c) (\(K_{\gamma} = -1 + \sqrt{1 + 2\gamma}\)) displays fast variations in \(\tilde{m}\), whereas positive \(\gamma\) clearly increases. For negative \(\gamma\) beyond \(|\gamma| > 0.5\), the (d) (\(K_{\gamma} = -1 - \sqrt{1 + 2\gamma}\)) likewise exhibits non-physical behaviour, with a consistent increase for positive \(\gamma\). Plots such as this illustrate the variation of magnetization with \(\gamma\) and different roots of \(K\), offering insights into the thermodynamic behavior of the system under different situations. Thus, the cases of (a) and (b) are shown quantum phase transitions. The important cases are (c) and (d) because they are determined the physical borders of the background.

To clarify the effect of quantum fluctuations on the system in the vicinity of certain values of absolute temperature, we will discuss quantum phase transitions in more detail by examining how the magnetization $\tilde{M}$ behaves close to the critical magnetic field values and four distinct scenarios that delineate distinct rotational parameter $K_{\gamma}$ regimes and the corresponding temperature impacts. Hence, Thus, Thus, the magnetization is determined for each case how the system reacts to changes in its external factors, such as temperature, rotational parameters, and magnetic field strength.

\textbf{Case 1: Behavior Near the Critical Point for Positive Real $K_\gamma$}

In this case, the magnetization $\tilde{M}$ is examined as the system gets closer to the critical magnetic field for positive real values of $K_\gamma$. According to which way $\gamma$ approaches zero, the study indicates different behaviors for $\tilde{M}$ when the magnetic field is near zero.

\begin{eqnarray}
   \lim_{\gamma\to 0^\mp}  \tilde{M}  &\simeq& \begin{cases} 
           -\frac{1}{2} + \frac{3}{2\tilde{\beta}} + \frac{\gamma}{6}\left(29 +\frac{55}{2\tilde{\beta}}\right),  & \gamma \rightarrow 0^- \\
          \frac{1}{2} - \frac{3}{2\tilde{\beta}} + \frac{5\gamma}{6}\left(1 + \frac{7}{10\tilde{\beta}}\right), & \gamma \rightarrow 0^+
       \end{cases}
\end{eqnarray}

For $\gamma$ $\rightarrow$ $0^{-}$  and $\gamma$ $\rightarrow$ $0^{+}$, the magnetization approaches - 1/2 and 1/2 at large inverse temperatures ($\tilde{\beta} \gg$ 1). This indicates a quantum phase shift in the system by showing symmetry breaking close to the critical magnetic field.

\textbf{Case 2: High-temperature regime Positive Real $K_\gamma$}

The system's magnetism diminishes with increasing temperature within the high-temperature domain $\tilde{\beta}$ $\rightarrow$ $0$. For both positive and negative values of $\gamma$ the magnetization is described by the following formulas as $\tilde{\beta}$ $\rightarrow$ $0$.
\begin{eqnarray}
   \lim_{\tilde{\beta}\to 0}  \tilde{M}  &\simeq& \begin{cases} 
           \sqrt{\frac{2}{\gamma}}\left[ 1 + \frac{\tilde{\beta}}{24}\left(77+\frac{1651\tilde{\beta}}{4}\right) \right],  & \gamma < 0 \\
          \sqrt{\frac{2}{\gamma}}\left[ 1 + \frac{\tilde{\beta}}{24}\left(77+\frac{811\tilde{\beta}}{4}\right) \right], & \gamma > 0
       \end{cases}
\end{eqnarray}

The magnetization gets closer to zero when the temperature rises, or when $\tilde{\beta}$ $\rightarrow$ $0$. This is due to the fact that at high temperatures, thermal effects reduce quantum effects because they prevail over the magnetic field.

\textbf{Case 3:Near the critical point for negative real $K_\gamma$}

For this particular case, the magnetization close to the critical magnetic field for real values of $K_\gamma$ that are negative is examined. Similar to Case 1, when $\gamma$ approaches zero from both directions, the system displays different behaviors for $\tilde{M}$.
\begin{eqnarray}
   \lim_{\gamma\to 0^\mp}  \tilde{M}  &\simeq& \begin{cases} 
           \frac{1}{2}- \frac{7}{8\tilde{\beta}} +\frac{\gamma}{2}\left(1-\frac{81}{32\tilde{\beta}}\right),  & \gamma \rightarrow 0^- \\
          -\frac{5}{2} + \frac{7}{8\tilde{\beta}} + \frac{5\gamma}{2}\left(1-\frac{81}{160\tilde{\beta}}\right), & \gamma \rightarrow 0^+
       \end{cases}
\end{eqnarray}
 It is noted that the magnetization approaches 1/2 in the case of $\gamma$ $\rightarrow$ $0^-$ and -5/2 in the case of $\gamma$$\rightarrow$ $0^+$ for large inverse temperatures, $\tilde{\beta}$ $\gg$ $0$. A significant discrepancy in $\tilde{M}$ behavior around the critical point between positive and negative magnetic fields indicates that the system is undergoing complex quantum phase transitions.

\textbf{Case 4:High-temperature regime for negative real $K_\gamma$}

Here, we investigate the magnetization behavior for negative real values of $K_\gamma$ in the high temperature range. Similar to Case 2, when temperature rises, the system's magnetism approaches zero. The formulas that follow define $\tilde{M}$ as $\tilde{\beta}$ $\rightarrow$ $0$.

\begin{eqnarray}
     \lim_{\tilde{\beta}\to 0}  \tilde{M}  &\simeq& \begin{cases} 
          - \sqrt{\frac{2}{\gamma}}\left[ 1 + \frac{\tilde{\beta}}{24}\left(77+\frac{1651\tilde{\beta}}{4}\right) \right],  & \gamma < 0 \\
          - \sqrt{\frac{2}{\gamma}}\left[ 1 + \frac{\tilde{\beta}}{24}\left(77+\frac{811\tilde{\beta}}{4}\right) \right], & \gamma > 0
       \end{cases}
\end{eqnarray}

The temperature increases are accompanied by a dramatic drop in magnetism. This demonstrates how the quantum effect is lessened at high temperatures. The system experiences phase transitions as a result of temperature changes and variations in the strength of the magnetic field, which are strongly influenced by quantum fluctuations and thermal processes.

$\tilde{M} \rightarrow 0$ since $\tilde{\beta}$ goes to zero much faster than $1/\sqrt{\gamma}$ when $\gamma \gg 1$. There is a second-order phase transition at real values of $K_\gamma$ near the critical value of the magnetic field at the low-temperature regime. $M = \begin{cases} 
          \mu_B,  & \gamma \rightarrow 0^- \\
          -\mu_B, & \gamma \rightarrow 0^+
       \end{cases}$ near the critical value of the magnetic field at $\tilde{\beta} = 1$ for the positive real $K_\gamma$ roots. $M$ = $-\mu_B$ near the critical value of the magnetic field at $\tilde{\beta} = 7/12$ for the negative real $K_\gamma$ values.

For $K_\gamma$ positive real values,
\begin{eqnarray}
    \begin{cases} 
           M = 0  & ,\tilde{\beta} = 3 \\
          \text{second-order phase transition}  & , \tilde{\beta} = \mathbb{R}_+ 
       \end{cases}
\end{eqnarray}

For $K_\gamma$ negative real values,  
\begin{eqnarray}
    \begin{cases} 
           \text{second-order phase transition}  & ,0 < \tilde{\beta} < 0.35 \\
            \text{first-order phase transition}  & ,0.35 < \tilde{\beta} < 1.75 \\
          \text{second-order phase transition}  & ,\tilde{\beta} > 1.75
       \end{cases}
\end{eqnarray}
thus highlights the diversity of quantum phase behavior displayed by the Dirac particle in the revolving, curved 2+1-dimensional spacetime. These numerous examples show how magnetization varies under various circumstances and add to our understanding of the system's critical points and thermodynamic characteristics.

\subsection{Heat Capacity}
This part provides a comprehensive explanation of the heat capacity's importance in phase transition research, building on the numerical analysis and theoretical foundation of the concept. Heat capacity is a crucial thermodynamic quantity to consider when determining if classical phase transitions take place in a system. It would offer details on the system's reaction to temperature variations. This formula can be used to define specific heat capacity
\begin{equation}
    \tilde{C} = \tilde{\beta}^2\frac{\partial^2 \ln Z}{\partial\tilde{\beta}^2},
    \label{Heat}
\end{equation}
where $\tilde{\beta}$ is the inverse temperature, which is obtained from $\tilde{\beta}$ = $\frac{m_e c^2}{k_B T}$, where T is the temperature, and $\tilde{C}$ is the dimensionless heat capacity, which is defined as the heat capacity divided by the Boltzmann constant $k_B$. This expression establishes a direct relationship between the system's heat storage capacity and the second derivative of the logarithm of the partition function Z with respect to $\tilde{\beta}$.

\begin{figure}[hbt!]
\centering
\begin{minipage}{.2\linewidth}
  \centering
  \includegraphics[width=\linewidth]{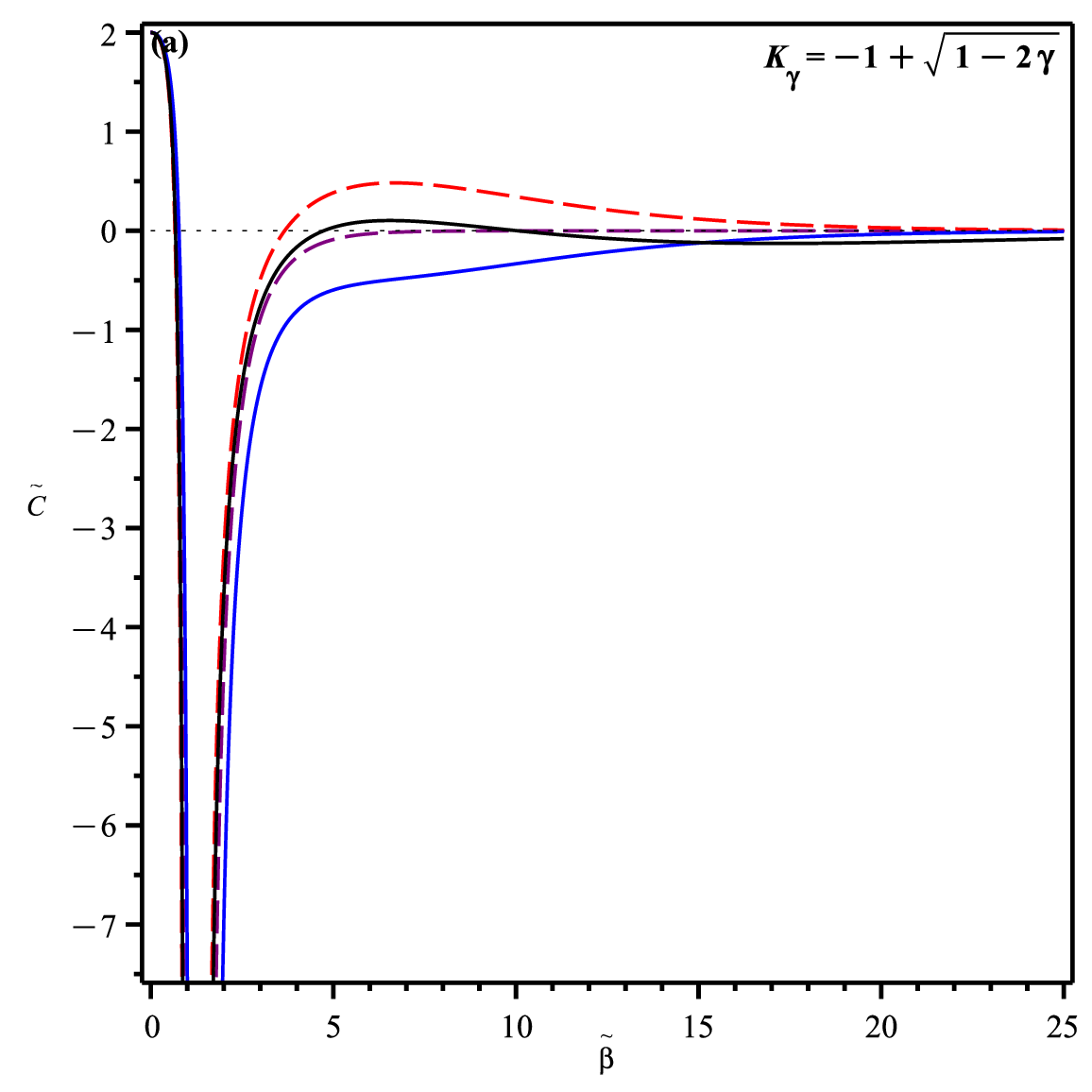}
  \label{CV1}
\end{minipage}
\quad
\begin{minipage}{.2\linewidth}
  \centering
  \includegraphics[width=\linewidth]{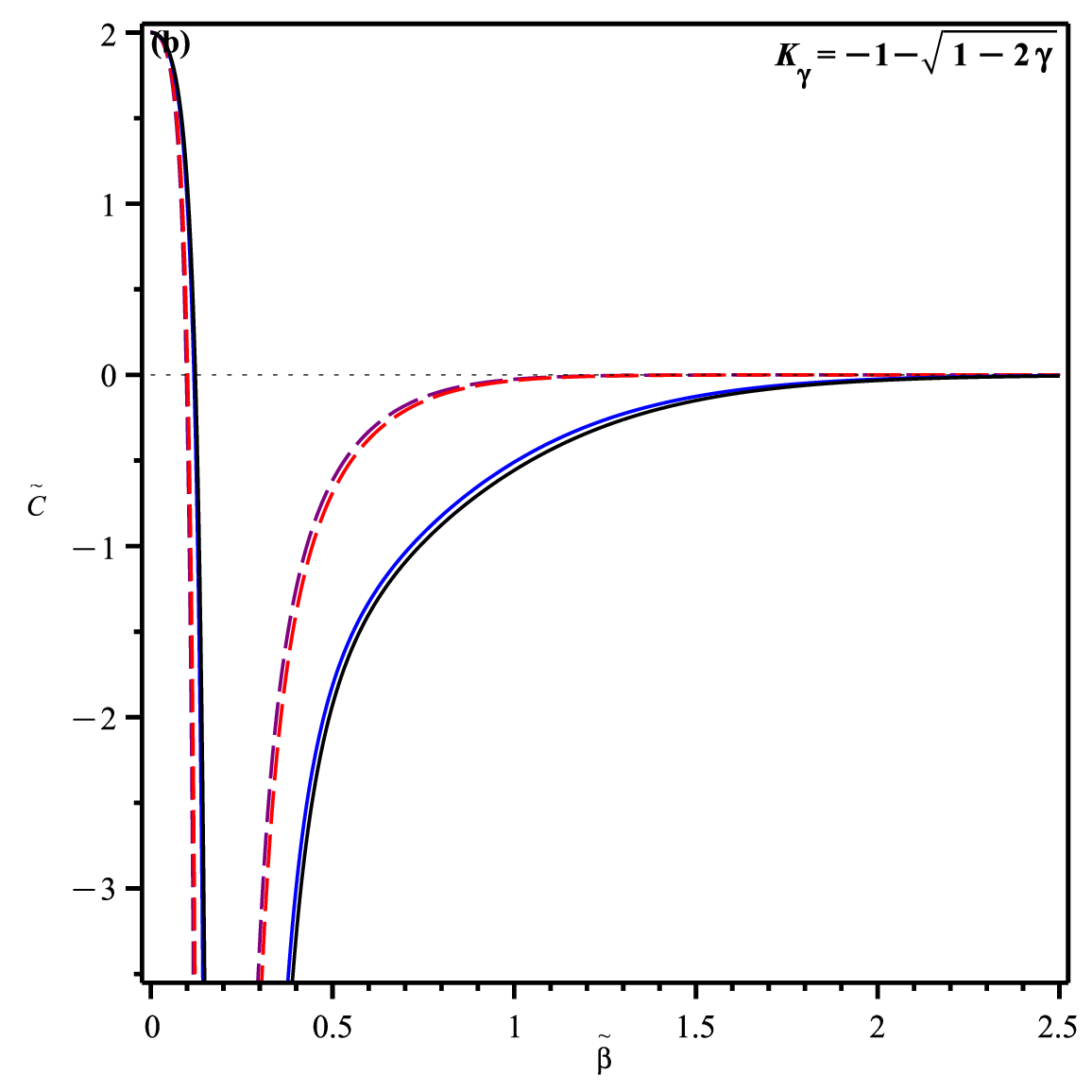}
  \label{CV2}
\end{minipage}
\begin{minipage}{.2\linewidth}
  \centering
  \includegraphics[width=\linewidth]{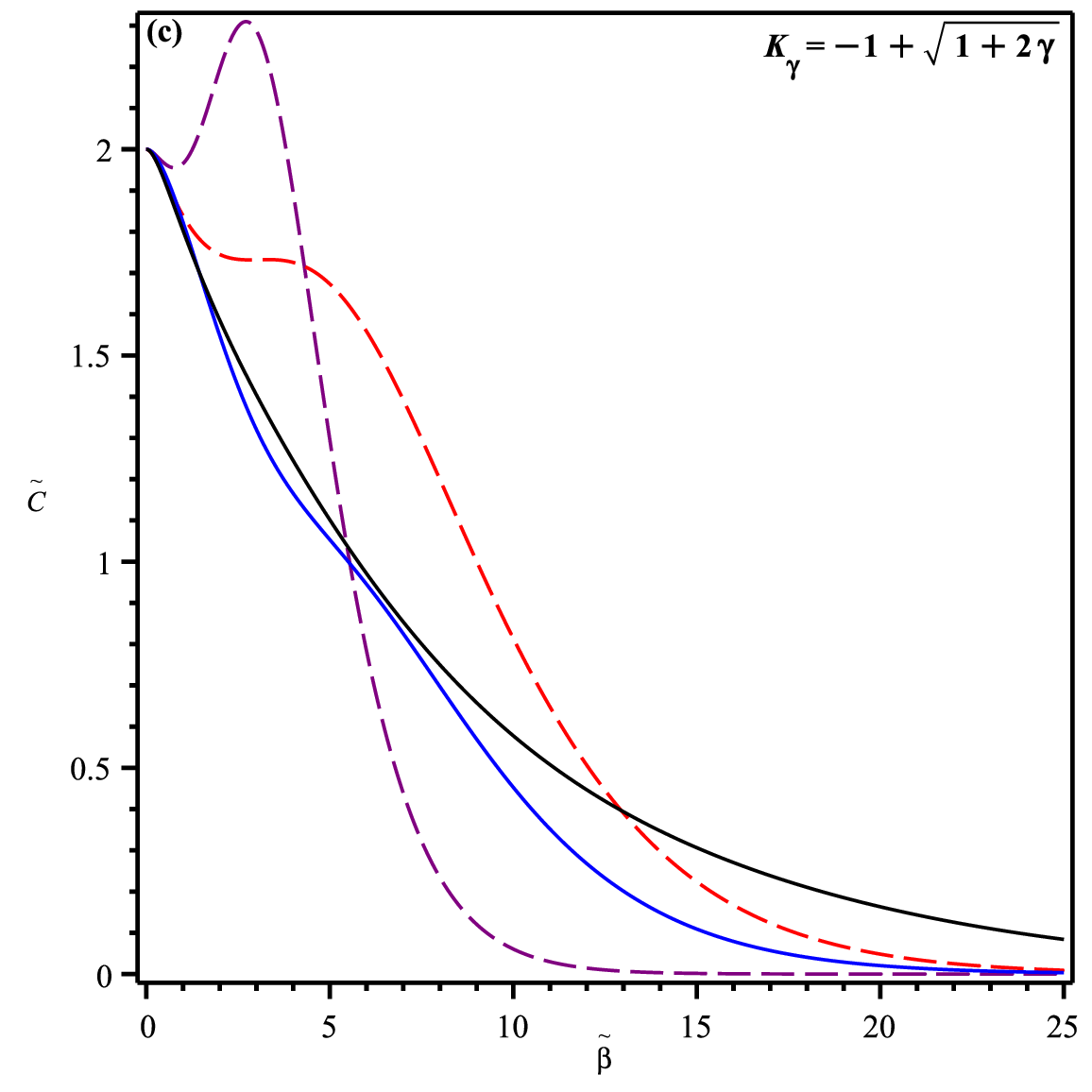}
  \label{CV3}
\end{minipage}
\quad
\begin{minipage}{.2\linewidth}
  \centering
  \includegraphics[width=\linewidth]{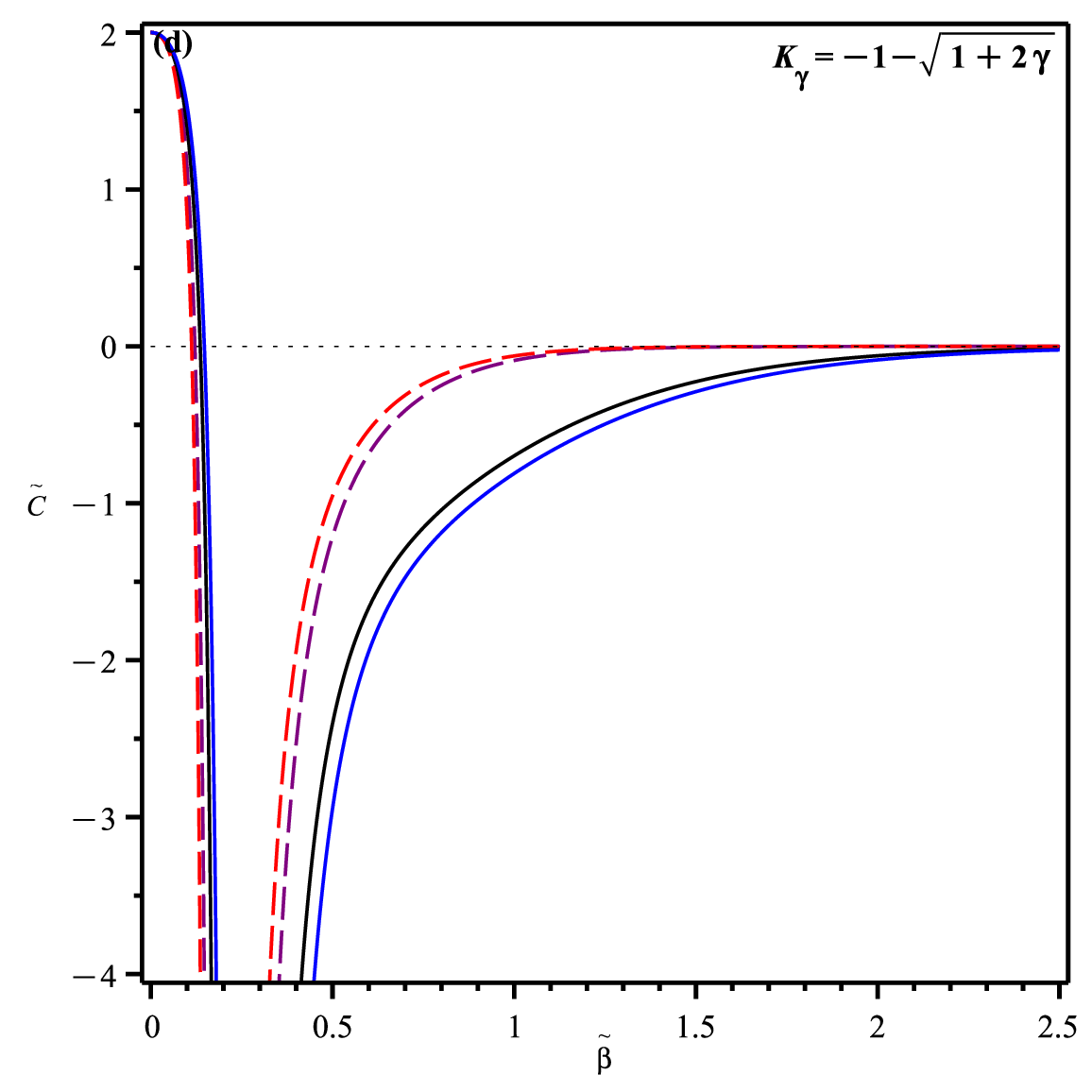}
  \label{CV4}
\end{minipage}
\caption{Plots of scaled heat capacity versus $\tilde{\beta}$ for all roots of K. The dashed $\left(\text{solid}\right)$ curves are LP $\left(\text{RP}\right)$ curves at $\gamma = -0.1, -0.2$ $\left(\gamma = 0.1, 0.2\right)$.}
\label{Fig2}
\end{figure}

The behavior of heat capacity in relation to various parameter values is very interesting, as Fig. \eqref{Fig2} illustrates. The numerical calculation of the specific heat, derived from the previous formula, is shown in the same graphic. Plots of the dimensionless heat capacity $\tilde{C}$ vs inverse temperature $\tilde{\beta}$ are shown for different roots of K. For various values of the magnetic field parameter $\gamma$, the dashed and solid lines, respectively, correspond to the left phase (LP) and right phase (RP) curves. Changes in the system's energy storage are indicated by $\tilde{C}$ typically tending toward negative values as $\tilde{\beta}$ rises.The behavior of $\tilde{C}$ exhibits notable variations due to the disparate roots of K and values of $\gamma$. For instance, before $\tilde{C}$ begins to converge, the plot at the (a), which corresponds to \(K_{\gamma} = -1 + \sqrt{1 - 2\gamma}\), shows an abrupt drop. This is in contrast to the plot, which decays more slowly, with \(K_{\gamma} = -1 - \sqrt{1 - 2\gamma}\), at the (b). While the (c) line for (\(K_{\gamma} = -1 + \sqrt{1 + 2\gamma}\)) has a maximum for an intermediate range of values of $\gamma$ and then rapidly goes to zero, the (d) graph for (\(K_{\gamma} = -1 - \sqrt{1 + 2\gamma}\)) shows a monotonically declining function of the heat capacity.

The relationship between temperature and magnetic field and heat capacity is seen in the following graphs. Depending on the value of $\gamma$, the system displays distinct thermodynamic characteristics for different roots. Phase transitions are indicated by peaks or abrupt decreases in the heat capacity.The fundamental understanding of the quantum and classical phase transitions in this system is based on the heat capacity analysis with respect to temperature and magnetic field. Distinct behavior for different roots and gamma shows how complex system thermodynamics are, and how important external influences like temperature and magnetic field strength are.
\section{\label{sec:level4}Summary and Conclusions}
This paper examines the dynamics of Dirac particles in a rotating, 2+1-dimensional curved spacetime that is uniformly magnetized, with a particular emphasis on quantum phase transitions. The present work explores the effects of rotating curved background geometry on the features of quantum systems subjected to external magnetic fields and quantum fluctuations by combining elements of quantum electrudynamics, general relativity, and statistical mechanics. The ground state of the system is altered by both thermal and quantum fluctuations when external variables, such as the strength of the magnetic field, are altered, in this framework, as well as by rotational parameters, as detailed by the research of quantum phase transitions.

Determining the energy spectra of the Dirac particles for the specified curved background is a significant result of this effort. Through the Dirac equation solution, the study clearly shows how the particle properties are affected by the externally provided magnetic field and the background's rotational parameter, $\Omega(r)$. These energy eigenvalues serve as the fundamental building blocks for any potential statistical mechanics and thermodynamics analysis of the system.

To enable the analysis of statistical mechanics and thermodynamics properties, the system's partition function was developed to make the shift from a microscopic model to a macroscopic one easier. Once the energy spectra were simplified, the partition function could be calculated. This provided important information about the several quantum phases that the system may have, depending on how the magnetic field was oriented and what energy levels were involved. The determination of the primary thermodynamic characteristics, including heat capacity and magnetization, was also made possible by examining the partition function.

In quantum phase transitions, magnetization is expressed as a dimensionless quantity. The observed result indicates phase transitions since magnetization exhibits a discontinuity close to the magnetic field's critical values. It was demonstrated that the magnetization behavior is dependent on the rotational parameter $k$ ; notable distinctions were noted between the magnetic field parameter $\gamma$'s positive and negative values. This variant demonstrates how closely spacetime geometry and quantum phenomena relate to each other in these systems.

Another thermodynamic variable that was investigated in the study was heat capacity, which showed how the system responded to temperature changes. The investigation revealed that the unique characteristics of the heat capacity were reliant on the rotational parameter's roots. Different conditions led to the appearance of first- and second-order phase transitions. These findings underlined the significance of temperature and magnetic field intensity as external variables on quantum phase transitions and provided additional evidence for the system's rich thermodynamic behavior.

All things considered, this study advances our understanding of quantum phase transitions in curved spacetime, particularly when it comes to Dirac particles affected by external magnetic fields. This work has important implications for theoretical physics, including applications to quantum electrodynamics and the investigation of the effects of curved spacetime on quantum systems. Subsequent researcher could delve deeper into these discoveries, examining the impact of diverse spacetime configurations or extraneous factors on quantum systems, potentially proposing novel perspectives for comprehending the quantum-classical boundary in intricate contexts.

\section*{Acknowledgment}
 We would like to thank Prof. Dr. Yusuf Sucu and Dr. Cavit Tekincay for valuable discussions and suggestions. E.S acknowledge the networking support of COST Actions CA21106-COSMIC WISPers in the Dark Universe: Theory, astrophysics and experiments (CosmicWISPers). We further thank  TUBITAK, SCOAP3, and ANKOS for their support.


\bibliography{aipsamp}

\end{document}